\documentclass[twocolumn,prb,showpacs,preprintnumbers,amsmath,amssymb,superscriptaddress]{revtex4}

\usepackage{graphicx}
\usepackage{dcolumn}
\usepackage{bm}

\begin{document}

\preprint{APS/123-QED}

\title{Comparison between experiment and calculated band structures for DyN and SmN}

\author{A.R.H. Preston}
\affiliation{MacDiarmid Institute for Advanced Materials and Nanotechnology, School of Chemical and Physical
Sciences, Victoria University of Wellington, Box 600, Wellington, New Zealand}

\author{S. Granville}
\affiliation{MacDiarmid Institute for Advanced Materials and Nanotechnology, School of Chemical and Physical
Sciences, Victoria University of Wellington, Box 600, Wellington, New Zealand}

\author{D.H. Housden}
\affiliation{MacDiarmid Institute for Advanced Materials and Nanotechnology, School of Chemical and Physical
Sciences, Victoria University of Wellington, Box 600, Wellington, New Zealand}
\author{B. Ludbrook}
\affiliation{MacDiarmid Institute for Advanced Materials and Nanotechnology, School of Chemical and Physical
Sciences, Victoria University of Wellington, Box 600, Wellington, New Zealand}

\author{B.J. Ruck}
\affiliation{MacDiarmid Institute for Advanced Materials and Nanotechnology, School of Chemical and Physical
Sciences, Victoria University of Wellington, Box 600, Wellington, New Zealand}

\author{H.J. Trodahl}
\affiliation{MacDiarmid Institute for Advanced Materials and Nanotechnology, School of Chemical and Physical
Sciences, Victoria University of Wellington, Box 600, Wellington, New Zealand}

\author{A. Bittar}
\affiliation{Industrial Research Ltd, Box 31-310, Lower Hutt, New Zealand}%

\author{G.V.M. Williams}
\affiliation{Industrial Research Ltd, Box 31-310, Lower Hutt, New Zealand}%

\author{J.E. Downes}
\affiliation{Department of Physics, Macquarie University, NSW 2109, Australia}

\author{K.E. Smith}
\affiliation{Department of Physics, Boston University, 590 Commonwealth Avenue, Boston, MA 02215, USA.}

\author{Y. Zhang}
\affiliation{Department of Physics, Boston University, 590 Commonwealth Avenue, Boston, MA 02215, USA.}

\author{A. DeMasi}
\affiliation{Department of Physics, Boston University, 590 Commonwealth Avenue, Boston, MA 02215, USA.}

\author{W.R.L. Lambrecht}
\affiliation{Department of Physics, Case Western Reserve University, Cleveland, Ohio 44106-7079, USA}

\date{\today}

\begin{abstract}
We investigate the electronic band structure of two of the rare-earth nitrides, DyN and SmN. Resistivity
measurements imply that both materials have a semiconducting ground state, and both show resistivity anomalies
coinciding with the magnetic transition, despite the different magnetic states in DyN and SmN. X-ray absorption
and emission measurements are in excellent agreement with densities of states obtained from LSDA+\textit{U} calculations, although for SmN the calculations predict a zero band gap.
\end{abstract}


\maketitle

\section{Introduction}

Despite several decades of research, many fundamental questions remain about the electronic structure of the
rare earth nitrides (RE-Ns). Both metallic and semiconducting states are found among early measurements, even
for materials of nominally the same composition.\cite{Hulliger,Vogt_Mattenberger} It has been only recently
that realistic band structure calculations have been performed, and these have heightened the interest by
predicting half metallicity in some members of the
series.\cite{Aerts_Svane,Larson_Schilfgaarde,Larson_Lambrecht,Duan_Tsymbal,Duan_Hardy,Antonov_Shpak,Chantis_Kotani,Ghosh_De,Johannes_Pickett} However, describing the
highly correlated 4\textit{f} electrons within band theory is a challenging problem, and as yet there is no
consensus amongst the predictions regarding the electronic states. Nevertheless, there is a strong desire to
explore these materials, and not only for the contribution to be made to the advancement of band-structure
calculation techniques; their properties as strongly spin-polarised conductors, either metallic or
semiconducting, can be expected to lead the way to improved materials for spintronics devices.

The most studied of the RE-Ns is
GdN,\cite{Duan_Hardy,Antonov_Shpak,Chantis_Kotani,Ghosh_De,Leuenberger_Hessler,Leuenberger_Wilhelm,Gerlach_Rauschenbach,Khazen_Ruck,Granville_Trodahl,Duan_Tsymbal,Larson_Lambrecht,Lambrecht}
which has an exactly half filled Gd 4\textit{f} shell and possesses the highest Curie temperature of the series.
Exchange splitting reduces the gap between the valence and conduction band states for the majority spin, but
various treatments disagree about whether the bands actually overlap to give a
half-metal\cite{Duan_Hardy,Aerts_Svane} or whether a gap persists.\cite{Larson_Lambrecht,Lambrecht} The first
attempt to calculate the properties of the full RE-N series, based on the local spin density approximation to
density functional theory (LSDA) plus a self-interaction correction, found electronic states ranging from
half-metallic to insulating.\cite{Aerts_Svane} More recently an approach based on LSDA plus Hubbard-\textit{U} corrections (LSDA+\textit{U}) has highlighted the importance of the 4\textit{f} electrons' orbital degrees of freedom in the cubic symmetry of the lattice.\cite{Larson_Schilfgaarde} There is an urgent need for quality experimental data addressing the band structure against which to compare predictions.

To a large extent the lack of reproducible experimental data is due to the propensity of these nitrides to
decompose in the presence of moisture in even very dry atmospheres.\cite{Hulliger,Gerlach_Rauschenbach}
Furthermore, it is difficult to ensure a low concentration of nitrogen
vacancies.\cite{Hulliger,Granville_Trodahl} It is thus important in this respect that recent advances have been
made toward the preparation and passivation of near-stoichiometric GdN
films.\cite{Granville_Trodahl,Gerlach_Rauschenbach} Conductance data on these very recent films have
established that GdN is a semiconductor at ambient temperature, and that despite a narrowing of the band gap it
remains so in the low-temperature ferromagnetic phase. There exist in addition x-ray absorption spectroscopy
results that show features in the conduction band (CB) density of states (DOS),\cite{Leuenberger_Hessler} and these results have recently been shown to be in good agreement with band structure calculations.\cite{Antonov_Shpak} There is a clear imperative to extend these results to a systematic study of the full band structure across the rare earth nitride series.

In the present work we describe an experimental study of nitrides of the RE elements with two fewer (Sm) and two more (Dy) 4\textit{f} electrons than Gd. We find excellent agreement between x-ray spectroscopy measurements and recent electronic structure calculations,\cite{Larson_Schilfgaarde} confirming the ability of the LSDA+\textit{U} method to describe key aspects of the band structures in these highly correlated materials. However, resistivity and magnetisation data show that both materials are semiconducting in their low-temperature magnetic ground states, in agreement with theory for DyN but implying that some fine-tuning of the LSDA+\textit{U} parameters is required for SmN. The understanding of these RE containing compounds and their behaviour in x-ray absorption and emission is of interest also in other highly correlated systems such as high-temperature superconductors or colossal magnetoresistance manganites.\cite{Kotliar_Marianetti,Kotani_Shin}

%
DyN, like GdN, is a ferromagnet with a Curie temperature ($T_{\mathrm{C}}$) reported variously as
17-26~K.\cite{Vogt_Mattenberger,Hulliger} SmN, in contrast, is reported to be antiferromagnetic with a N\'{e}el
temperature ($T_{\mathrm{N}}$) of 15-18~K,\cite{Vogt_Mattenberger,Hulliger} although it should be noted that
the moment per Sm ion is very small and the antiferromagnetic assignment has not been confirmed. Recent
calculations assuming a ferromagnetic ground state have found that DyN is a semiconductor while SmN is a half
metal.\cite{Aerts_Svane,Larson_Schilfgaarde} The calculated gaps in the paramagnetic phase quoted in
Ref.~[\onlinecite{Larson_Schilfgaarde}] are in reasonable agreement with early optical
measurements.\cite{Hulliger} The conductivity of these early samples of SmN and DyN was reported to show
semi-metallic behaviour, although it is certain that these are influenced by imperfect stoichiometry and
impurities. There have been no recent reports of the conductivity of SmN or DyN.

\section{Sample Growth and Characterisation}

We have grown films of these two nitrides using the technique developed for GdN, by vacuum evaporation of the RE
metals in the presence of a scrupulously pure atmosphere of N$_2$ at a pressure of
$10^{-4}$~mbar.\cite{Granville_Trodahl} \textit{In situ} resistance measurements were performed to a
minimum temperature of 100~K, and \textit{ex situ} studies performed on films passivated by a layer of
nano-crystalline GaN. Consistency between the \textit{in situ} and \textit{ex situ} resistivity results confirms
the effectiveness of the capping layers. Rutherford backscattering spectroscopy has established that the films
have the correct stoichiometry to within the detection limit of about 2\%, and their rock salt structure has
been shown by x-ray diffraction to consist of typically 8~nm crystallites with the expected lattice constants.

\section{Results and Discussion}

\begin{figure}
\includegraphics[width=8cm]{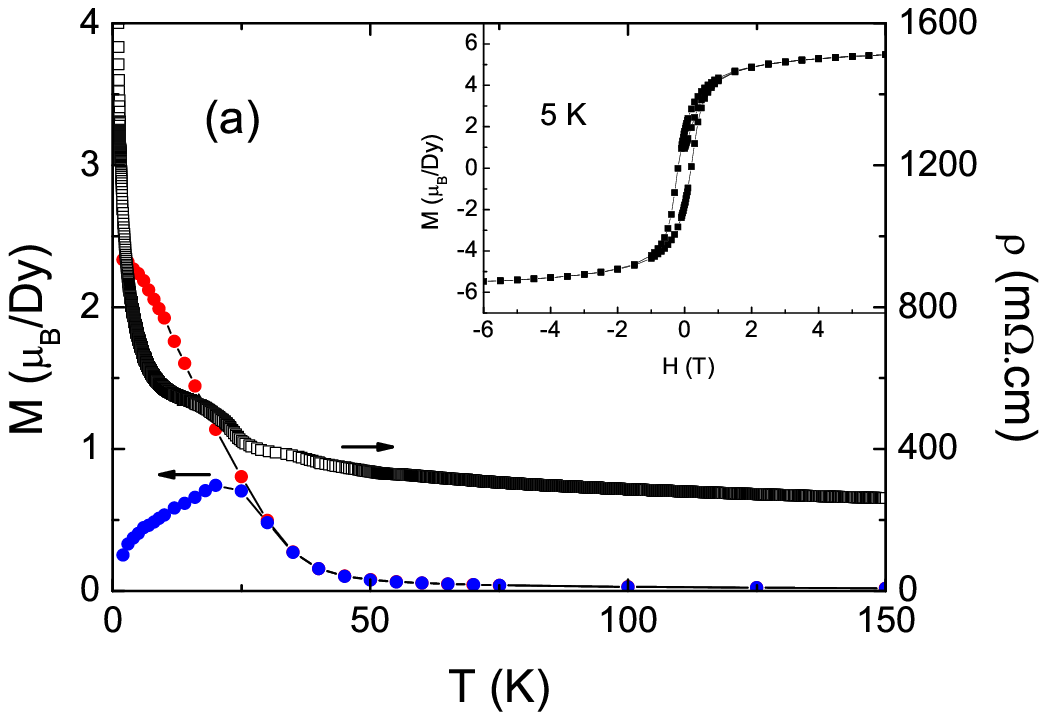}
\includegraphics[width=8cm]{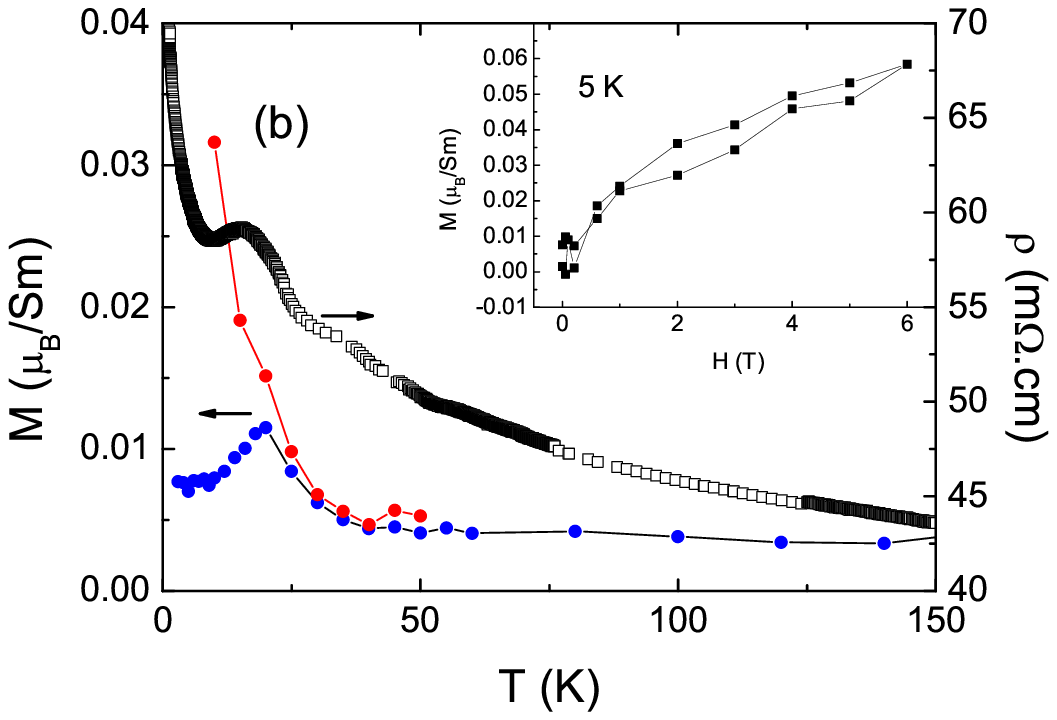}
\caption{\label{MandR}(color online) Magnetisation and resistivity of (a)~DyN, and (b)~SmN (insets show magnetisation versus field at 5~K). Blue symbols show zero-field cooled magnetisation data, red symbols show field-cooled data.}
\end{figure}

Figures~\ref{MandR}(a) and (b) show the magnetisation and resistivity of, respectively, DyN and SmN. Turning
first to the magnetisation, DyN shows a divergence between the field-cooled and zero-field-cooled data below
25~K. Field sweeps at 5~K show a hysteresis and saturation (see inset), making it clear that the low temperature
phase is ferromagnetic, with the T$_{C}$ of around 25~K near the top of the range reported in the
literature.\cite{Hulliger,Vogt_Mattenberger} The response of SmN is much weaker at all temperatures, despite
the use of a measurement field of 5000~Oe rather than the 500~Oe used for DyN, and there is an approximately
constant offset of about 0.004$\mu_{B}$ per Sm associated with uncertainties in removing the background signal
from the Si substrate. Nevertheless, the field-cooled and zero-field-cooled data separate implying a magnetic
transition at around 20~K, slightly higher than previously reported transition
temperatures.\cite{Hulliger,Vogt_Mattenberger} No clear hysteresis is seen at 5~K, so it is likely that SmN is
in fact metamagnetic (i.e., an antiferromagnet in which the moments can be ferromagnetically aligned by an
applied field). Even in a field of 6~T the low temperature moment is less than 0.1$\mu_{B}$ per Sm, considerably
less than the theoretical saturated moment for atomic Sm ions of $gJ\mu_B=0.71\mu_B$. This is consistent with
the calculations of Ref.~[\onlinecite{Larson_Schilfgaarde}] which emphasize the importance of orbital magnetic
moment in these materials and for SmN in a ferromagnetic state essentially predict a cancellation of the orbital
and spin moment, i.e., $L_z+2S_z=0$.

The magnitude of the resistivity is a factor of ten greater in DyN than in SmN and the temperature dependence
more pronounced, although for both materials the behaviour is typical of a heavily doped semiconductor and any
activation energy extracted from the data is orders of magnitude smaller than the calculated band gaps.
Nevertheless the trends are consistent with the calculated gaps if the energy of the defect states scales with
the band gap. Anomalies are seen at the magnetic ordering temperature, strongly reminiscent of the behaviour of
GdN in which exchange splitting of the bands causes a reduction in the relevant excitation
energy.\cite{Leuenberger_Hessler,Granville_Trodahl} These measurements are taken at zero field, so in light of
the previous discussion it is interesting that the anomaly exists in SmN as well as in ferromagnetic DyN. In any
case, the temperature dependence of the resistivity establishes that both materials are semiconducting over the
entire temperature range. We now discuss measurements that address the band structure directly.

We have performed x-ray absorption near-edge spectroscopy (XAS) at the N K-edge to give a direct measure of the
N~\textit{p}-projected empty-state partial DOS (PDOS), complemented with x-ray emission spectroscopy (XES) to
plot out the filled-state N~\textit{p} PDOS. For the x-ray spectroscopy the growth was carried out in the
preparation chamber on the X1B synchrotron beamline at the National Synchrotron Light Source at Brookhaven
National Laboratory, and measurements completed \textit{in situ} without breaking the vacuum. X-ray absorption
spectra were recorded using the total electron yield, with the spherical grating monochromator
providing an energy resolution of 0.2~eV. X-ray emission spectra were obtained using a Nordgren-type
grazing-incidence soft x-ray spectrometer with a resolution of 0.4~eV. Both XAS and XES were recorded at room
temperature.

\begin{figure}
\includegraphics[width=7.5cm]{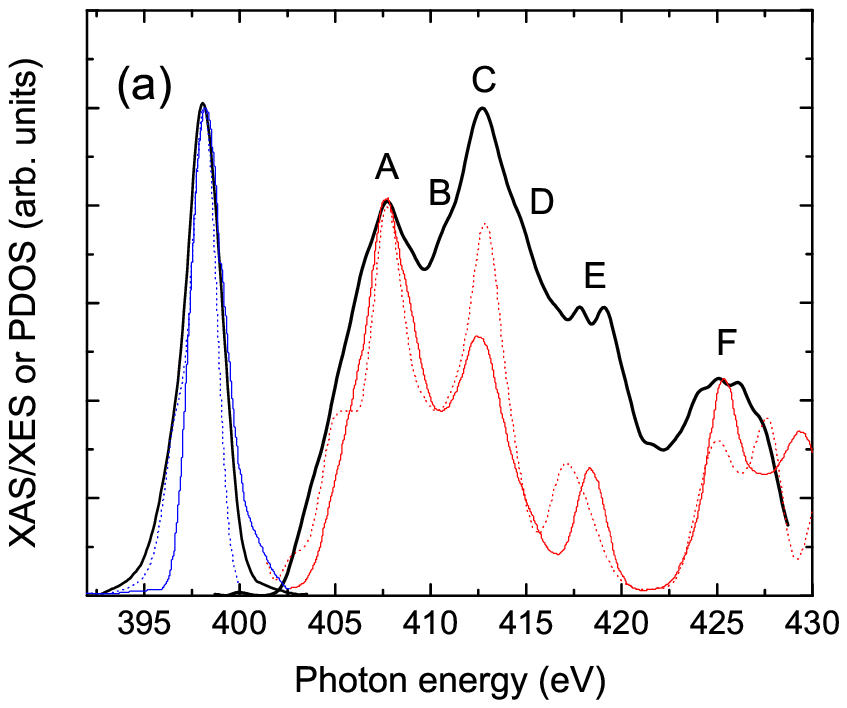}
\includegraphics[width=7.5cm]{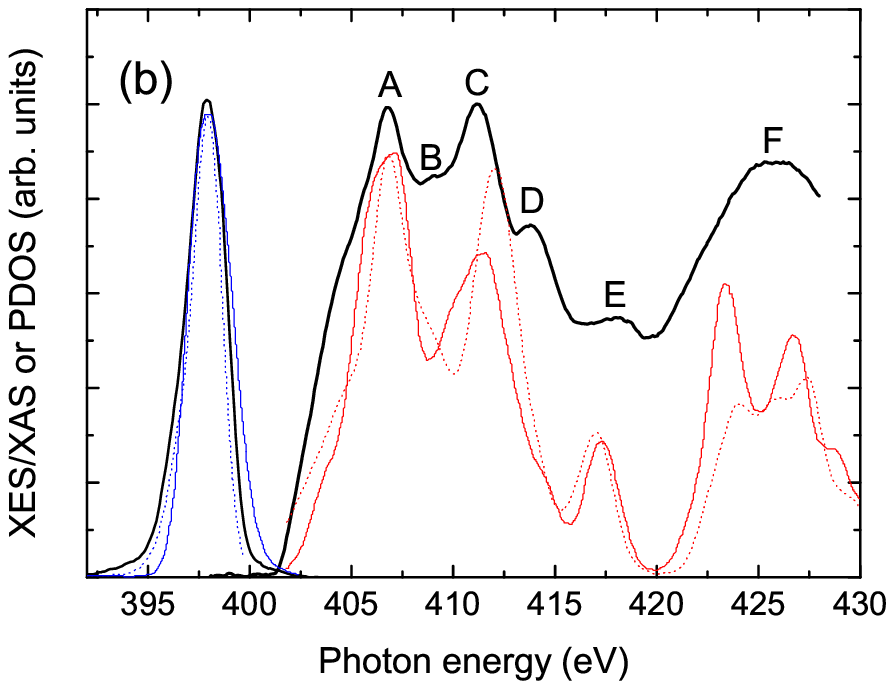}
\includegraphics[width=7.5cm]{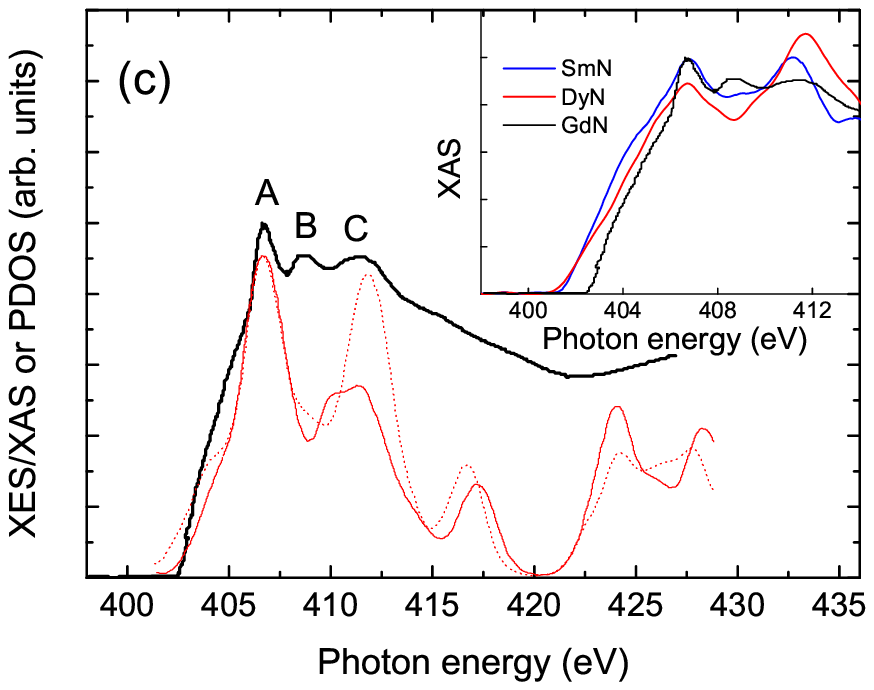}
\caption{\label{XASXES}(color online) Measured x-ray absorption (thick black line) and emission (thin black line) for (a)~DyN, (b)~SmN, and (c)~GdN,~\cite{Leuenberger_Hessler} compared to the calculated nitrogen PDOS. The calculations are performed both with (solid lines) and without (dotted lines) including the core hole. The inset to~(c) compares the measured absorption for all three materials.}
\end{figure}

In Figure~\ref{XASXES} the XAS/XES measurements of the PDOS are displayed, normalised separately to their
maximum values. The absolute energies are somewhat arbitrary, but the relative energy scales have been carefully
calibrated using the elastic peaks in XES spectra obtained at different excitation energies. Note that, apart
from the location of the elastic peak, the XES spectra were insensitive to the excitation energy. The
calibration was further confirmed using the known L-edge peak positions of a reference Co sample.

The XES looks qualitatively similar in DyN and SmN with both displaying a relatively narrow asymmetric band. The
absorption spectra of both materials also exhibit strong similarities, with a broad rise above the absorption
edge followed by a series of peaks stretching more than 25~eV. Both DyN and SmN show a clear energy gap between
the absorption and emission, which we will discuss in detail below. To further the comparison XAS for GdN
obtained from Ref.~[\onlinecite{Leuenberger_Hessler}] is also included in the plots. Once again the same set of
principal features are present but with slightly lower resolution, presumably due to the use of a Cr capping
layer on the GdN.

A comparison with XAS data from a thin film GaN sample containing molecular nitrogen\cite{Ruck_Markwitz} showed
that the $\pi^{*}$ resonance of N$_2$ appears in the same energy range as peak A, although these features in our
RE-N spectra are much too broad to assign solely to the molecular signal. On the other hand a clear N$_2$ signal
appears in the XAS after prolonged beam exposure, and a similar effect was noted for the GdN in
Ref.~[\onlinecite{Leuenberger_Hessler}] for which peak A is narrower than those in DyN or SmN. Thus, the XAS
peak A is intrinsic to the RE-N samples, but some slight interference from molecular N$_2$ trapped within the
films during growth may also be present.


To extract information about the band structure we include in Fig.~\ref{XASXES} the calculated PDOS for all three materials obtained from detailed band structure calculations. The PDOS were calculated using the full-potential linearized muffin-tin orbital method\cite{Methfessel_Casali} (FP-LMTO) within the LSDA+\textit{U} approach.\cite{Liechtenstein_Zaanen,Larson_Lambrecht,Larson_Schilfgaarde} The Hubbard-\textit{U} correction applied to the RE 4\textit{f} states is based on previous BIS/XPS studies of Gd compounds, and the expected scaling of the screened $U_f$ with the bare Coulomb interactions in the other REs taken from atomic calculations. The trend in $U_f$ values was discussed in Ref.~[\onlinecite{Larson_Schilfgaarde}]. An additional shift is applied to the RE 5\textit{d} states, although here $U_d$ is used to correct essentially different physics, namely, the usual underestimate of the gap in semiconductors. In the present materials, this leads to a shift in RE-\textit{d} states. The $U_d$ value for GdN was recently revisited based on a new study of the optical properties and submitted as a separate paper,\cite{Trodahl_GdN} but we note that the revised $U_d$ does not alter significantly the shape of the conduction band N \textit{p}-like PDOS or the comparison to XAS.  For SmN and DyN optical data are not yet available, and thus the same $U_d$ value was used as in Ref.~[\onlinecite{Larson_Schilfgaarde}]. For comparison to the room temperature experimental data the calculated PDOS have been summed over the two spin orientations.


The measured gap between the XAS and XES is about 1.5~eV for DyN and SmN, about 1~eV larger than the calculated
values quoted above. However, XAS creates a core hole that shifts the measured spectra by uncertain amounts
dependent on the degree of screening and the projection of the states onto the core hole. In fact, analysis of the dynamical many-body effects in model systems suggests that the correct potential to use for the absorption includes the core hole on the absorbing atom, while for emission the correct potential is that without the core hole\cite{vonBarth_Grossmann,Mahan} (the ``final-state rule"). Others have found that the best description is achieved somewhere between the initial- and final-state rules and have suggested that it depends on whether the system is metallic or insulating.\cite{Luitz} Calculations for both potentials are included in Fig.~\ref{XASXES}, where the core hole case was calculated using a simple cubic cell with a basis in which every fourth nitrogen atom has a core hole. In light of this we do not attempt to compare the measured and calculated gaps, but instead adjust the energy axis of the calculated filled and empty states separately to best match the experiment.

The calculation with no core hole gives an excellent description of the valence band N \textit{p} PDOS, including the asymmetry on the low energy side. In contrast note the presence of a shoulder on the high energy side of the calculation that includes the core hole which is clearly not present in the experiment. Thus we can confirm the accuracy of the band structure calculation and the obeyance of the final state rule for the valence band. Note that the calculation predicts a small contribution to the bottom of the band associated with hybridisation between nitrogen 2\textit{p} states and the highest occupied RE 4\textit{f} states. There is spectral weight at the bottom of the XES which may be attributed to this, although with the present experimental resolution it is difficult to be certain. A comparison with XES data from GdN, which lacks such states, would be interesting in this respect.

For the XAS, the core hole calculation provides a remarkably good fit, reproducing even the high-energy features
E and F in DyN and SmN. There are also shoulders hinting at peak B for GdN and the weak features B and D for
SmN. This provides strong evidence that the present band structure calculation correctly describes the unfilled
density of states for these RE-nitrides. It should be noted that the calculation with no core hole also performs
well, and looks quite similar to the core hole calculation, in contrast with recent results obtained for  GaN.\cite{Strocov_Nilsson} This is largely due to the fact that the N \textit{p} PDOS in the conduction band can be interpreted rather as the \textit{p}-component of a partial wave
expansion inside the N sphere of the tails of the predominantly RE \textit{d}-like and some RE \textit{f}-like
atomic states extending into the N spheres. Thus, they are not as sensitive to the perturbation of the core hole
as the XES. This explanation is based on the LMTO terminology.\cite{Andersen,Andersen_Jepsen} Alternatively, in a linear combination of atomic orbitals point of view, the conduction bands are antibonding combinations of RE \textit{d} (and RE \textit{f}) orbitals with N \textit{p} orbitals. Their energy difference implies that the higher states have only a small N \textit{p} contribution and it is precisely this contribution we are measuring.

To further investigate systematic variations across the RE-nitride series the absorption edges of all three
materials are compared in the inset to Fig.~\ref{XASXES}(c). Here, the energy axes have been adjusted to align
the peaks A, which are calculated to lie at a similar energy above the Fermi level in each case. There are
differences between the peak locations and relative peak strengths, which should form a useful guide for any
subsequent refinement of the band structure calculations. The absorption onset for GdN is also pushed up in
energy relative to DyN and SmN. This can be interpreted as evidence of unoccupied bands of low-dispersion and
high RE \textit{f} character just above the Fermi level in SmN and DyN, that provide the major contribution to
the N \textit{p} PDOS in this energy region. Such states are not expected for GdN.\cite{Larson_Schilfgaarde,Antonov_Shpak}

\section{Conclusions}

In summary, we have provided much needed experimental data addressing the band structure of two of the
rare-earth nitrides, namely DyN and SmN. Our x-ray spectroscopy data confirm the ability of LSDA+\textit{U}
calculations to reproduce the main features of their band structures, although some fine tuning
may be required to reproduce the band gaps implied by resistivity data. It
should be noted in this context that the calculated gap depends sensitively on the $U_d$ parameter
which controls the position of the RE-\textit{d} bands and was adjusted in the calculations to the optical
absorption threshold in GdN reported by Hulliger \textit{et al.}~\cite{Hulliger} and identified with the average
of spin-up and spin down direct transitions at the $X$-point. Experimental studies of the optical absorption
threshold for DyN and SmN are planned and may assist in the determination of this $U_d$ parameter in the
LSDA+\textit{U} theory.

We note also the contribution that might be made by x-ray photoelectron spectroscopy (XPS) or Brehmsstrahlung isochromat spectroscopy (BIS) to understanding the band structure of these materials. These techniques probe the total density of states, and have proven useful for investigating the location of the 4\textit{f} levels in various gadolinium pnictides,\cite{Ghosh_De} but to our knowledge no data presently exist for SmN or DyN.

\section{Acknowledgements}

We thank the Tertiary Education Commission of NZ for support through the Centres of Research Excellence fund.
DHH thanks the NZ Science, Mathematics and Technology Teacher Fellowship scheme, administered by the Royal
Society of NZ. The work at Case Western Reserve University was funded by the U.S.~Army Office of Research under
grant number W911NF-06-1-0476. The Boston University program is supported by the NSF under grant number DMR-0311792.  The NSLS is supported by the U.S.~Department of Energy, Division of Materials and Chemical Sciences.



\end{document}